# Two-dimensional nanovaristors at grain boundaries account for memristive switching in polycrystalline BiFeO$_3$

Xiao Shen,[1,*,†] Kuibo Yin,[2,3,*,†] Yevgeniy S. Puzyrev,[1,*,†] Yiwei Liu,[4] Litao Sun,[2,†] Run-Wei Li,[4,†] and Sokrates T. Pantelides[1,3,5]

[1] Department of Physics and Astronomy, Vanderbilt University, Nashville, TN 37235

[2] SEU-FEI Nano-Pico Center, Key Laboratory of MEMS of Ministry of Education, Southeast University, Nanjing 210096, China

[3] Materials Science and Technology Division, Oak Ridge National Laboratory, Oak Ridge, TN 37831

[4] Key Laboratory of Magnetic Materials and Devices, Ningbo Institute of Materials Technology and Engineering, Chinese Academy of Sciences, Ningbo, Zhejiang 315201, China

[5] Department of Electrical Engineering and Computer Science, Vanderbilt University, Nashville, TN 37235

---

[*] These authors contributed equally to this work
[†] Email: xiao.shen@vanderbilt.edu (X.S); yinkuibo@seu.edu.cn (K.Y.); yevgeniy.s.puzyrev@vanderbilt.edu (Y.S.P.); slt@seu.edu.cn (L.S.); runweili@nimte.ac.cn (R.-W.L.)

**Graphical Abstract**

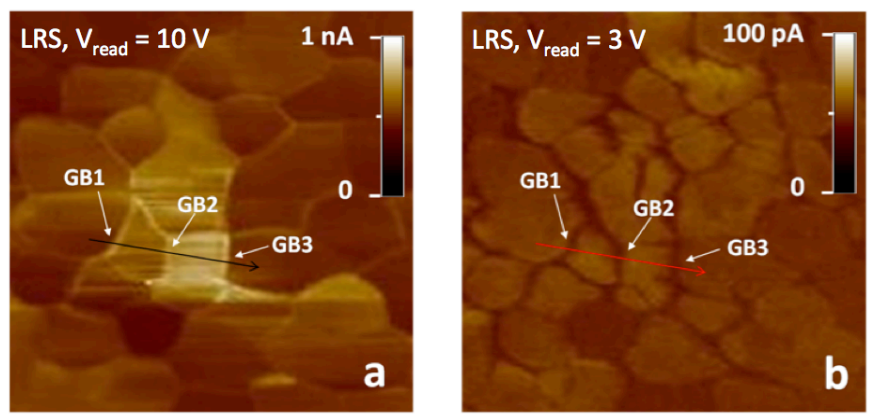

**Abstract:**

Memristive switching in polycrystalline materials is widely attributed to the formation and rupture of conducting filaments, believed to be mediated by oxygen-vacancy redistribution. The underlying atomic-scale processes are still unknown, however, which limits device modeling and design. Here we combine experimental data with multi-scale calculations to elucidate the entire atomic-scale cycle in undoped polycrystalline $BiFeO_3$. Conductive atomic force microscopy reveals that the grain boundaries behave like two-dimensional nanovaristors while, on the return part of the cycle, the decreasing current is through the grains. Using density-functional-theory and Monte-Carlo calculations we deduce the atomic-scale mechanism of the observed phenomena. Oxygen vacancies in non-equilibrium concentrations are initially distributed relatively uniformly, but they are swept into the grain boundaries by an increasing voltage. A critical voltage, the SET voltage, then eliminates the barrier for hopping conduction through vacancy energy levels in grain boundaries. On the return part of the cycle, the grain boundaries are again non-conductive, but the grains show nonzero conductivity by virtue of remote doping by oxygen vacancies. The RESET voltage amounts to a heat pulse that redistributes the vacancies. The realization that nanovaristors are at the heart of memristive switching in polycrystalline materials may open possibilities for novel devices and circuits.

Metal oxides have been widely studied for their intrinsic physical properties, such as colossal magnetoresistance [1], multiferroicity [2] and, more recently, memristive switching [3, 4, 5], which opens new possibilities to use these materials in resistive random access memories (RRAM), [6] analog devices, [7] and neuromorphic circuits [8]. In the case of undoped polycrystalline oxides, memristive switching has been attributed to the formation and destruction of conductive channels [9, 10]. It is often assumed that these channels are metallic filaments [11, 12]. Redistribution of oxygen vacancies has been frequently invoked as the origin of the memristive phenomenon, but their actual role cannot be deduced from the data.

Memristive switching has been observed in $BiFeO_3$ (BFO), a metal oxide studied mainly for its multiferroic properties [2, 13-15]. The memristive properties of BFO are rather diverse and depend on the crystallinity and doping [16-21]. The switching of single crystal [16] and heavily La-doped polycrystalline BFO [17] exhibits an asymmetric I-V curve, which has been attributed to the change of ferroelectric polarization. The switching of undoped [18] and heavily Nd-doped polycrystalline BFO [19] exhibits a symmetric I-V curve. For the undoped polycrystalline BFO, local measurements of conductivity were recently performed using conducting atomic force microscopy (c-AFM) and redistribution of oxygen vacancies in grain boundaries was suggested to be the origin of memristive switching [18]. The experimental data, however, were limited to a single probing voltage and the precise role of oxygen vacancies could not be deduced.

In this paper, we combine a comprehensive characterization of local conduction channels in undoped polycrystalline $BiFeO_3$ at different probing voltages with pertinent multi-scale theoretical calculations and construct a detailed atomic-scale account of the observed memristive behavior. We show that, contrary to the prevalent metallic filament basis for memristive switching, grain boundaries in BFO behave as switchable two-dimensional nanovaristors. We first present c-AFM data showing essentially zero conduction until a critical voltage is reached (the SET voltage), which is tantamount to a varistor behavior. On the return part of the hysteresis loop, the current decreases gradually, but conduction is only through the grains, confirming the varistor nature of the grain boundaries. Calculations show that, under the increasing voltage, oxygen vacancies leave the grains and enter grain boundaries, as recently demonstrated in the case of ZnO [22]. There is no conductive path, however, until a critical voltage, which we identify as the SET voltage, eliminates the energy barrier for hopping conduction through oxygen-vacancy levels in the energy gap. On the return path, conduction only through the grains is consistent with the predicted varistor behavior of grain boundaries and the fact that vacancy aggregation in the grain boundaries amounts to both a removal of Coulomb scattering centers from the grains and remote doping by the vacancies in the grain boundaries. Finally, the RESET process by a negative voltage can be attributed to a cooperative effect in which charged oxygen vacancies in the grain boundaries boost the effective local voltage into a pulse that produces local heating and re-disperses the vacancies into the grains.

$BiFeO_3$ films were prepared on commercial $Pt/Ti/SiO_2/Si$ substrates by a sol-gel method and annealed at 700 °C for 10 min in air as in Ref. 18. The composition of the films was examined by X-ray diffraction (D8 Advance, Bruker AXS) and no impurity phase was detected. Cross-sectional TEM (Tecnai F20, FEI) results indicated that the film thicknesses were about 210 nm and the average grain size was about 100 nm. The atomic force microscopy (Dimension, Veeco) results showed that the film surface was smooth with a mean square roughness of 15 nm.

The memristive switching behavior of the BFO films was probed by the c-AFM method. The conducting tip of the c-AFM was grounded and dc voltage was applied to the Pt bottom electrode. Figure 1 shows the morphology of film (Fig. 1a) measured by AFM. The current mappings with different dc voltages in the same scanning region were obtained simultaneously (Fig. 1b-1g). The external bias voltage was changed in a sequence of 3 V → 6V → 10 V → 3 V → -5V → 3 V, and the current maps shown in the figures were read at those voltages. At 3 V (Fig. 1b), there is no measurable current and the film is at high resistive state (HRS). At 6 V, there is significant current through the film, more than 500 pA, and the grain boundaries are quite dim (Fig. 1c). In contrast, a larger voltage of 10V SETs the film to a truly low resistive state (LRS), in which the current through grain boundaries is ~700 pA (Fig. 1d). Lowering the reading voltage to 3 V, there is still a current of ~50 pA (Fig.1e), but it clearly flows through the grains – the grain boundaries are noticeably dark. A negative voltage of -5V (Fig. 1f) RESETs the film back to HRS and no current is observed at 3 V reading voltage (Fig. 1g). These c-AFM results clearly show memristive switching of undoped BFO, which is consistent with earlier work [18]. Furthermore, the new results provide us with information to extract the precise atomic-scale mechanism that underlies the switching process.

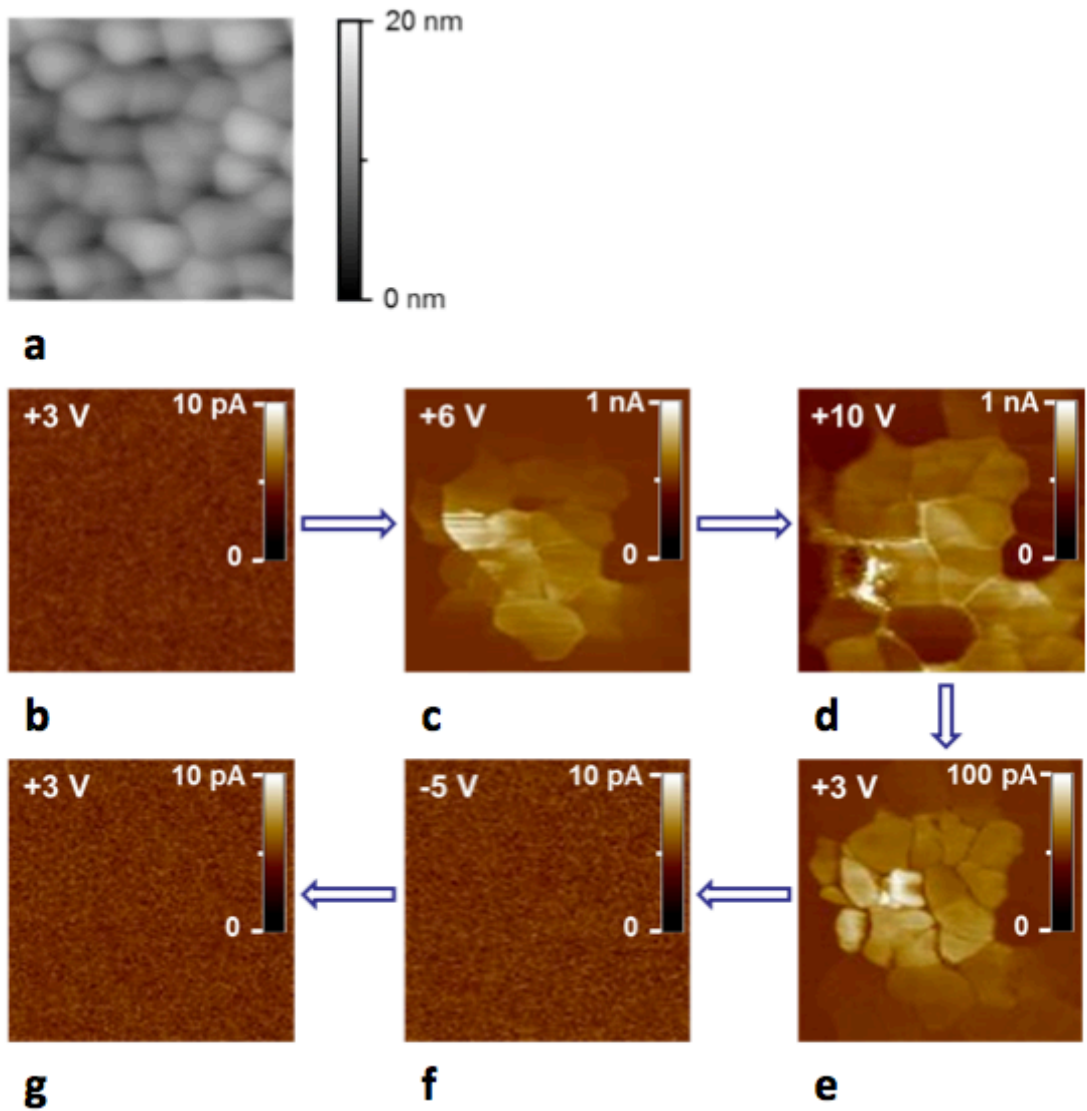

**Figure 1. Current images at several voltages during a complete cycle. a**, AFM image shows the morphology of the sample with a scanning size of 550 × 550 nm$^2$. **b-g**. c-AFM images of the current signal with different external bias during a complete cycle.

In order to confirm the role of the grains and the grain boundaries in the memristive cycle in the BiFeO$_3$ film, another area of the film was SET at 10V and then read at 3 V. The results shown in Figure 2 indicate that the sample is successfully SET to LRS by the writing process at 10 V.

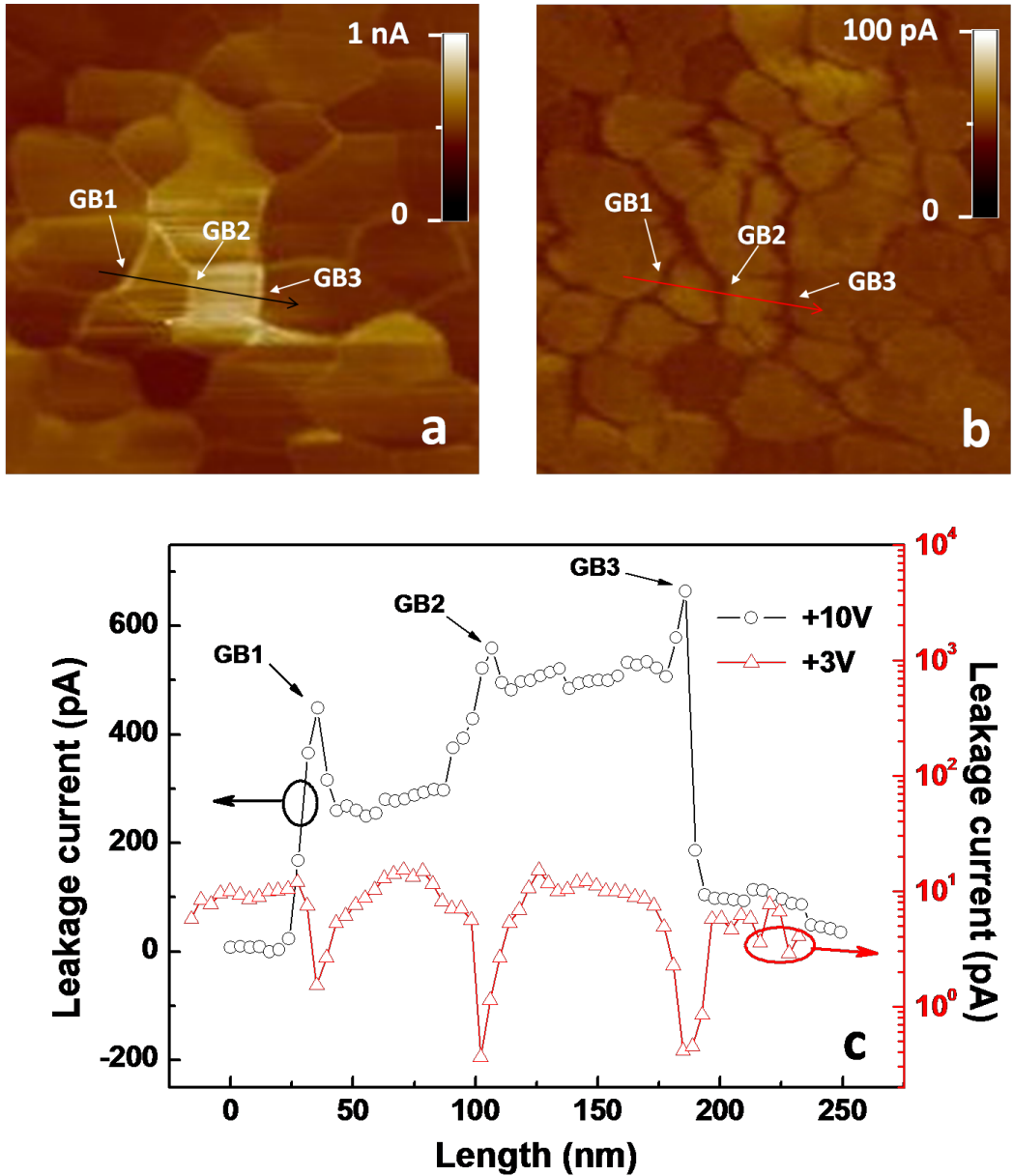

**Figure 2. Current images through grains and grain boundaries**. **a**, Current signal image with a +10 V bias voltage applied to the bottom electrode. The arrow length is 250 nm. **b**, Current signal image with a +3 V bias voltage applied to the bottom electrode after the process **a**. **c**, Current signals along the arrow directions in **a** and **b**, respectively.

The contrast of grain boundaries at 10 V (Fig. 2a) and 3 V (Fig. 2b) is confirmed from the images. A line scan unambiguously shows the enhancement of the current at the location of 3 grain boundaries, which are marked as GB1, GB2 and GB3 in Fig. 2a and 2b. Figure 2c shows the current as a function of position along the line at both 10 V and 3 V reading. It is clear that the grain boundaries are less conducting than the grains at 3 V, as the current profile has three deep dips at their locations; but become highly conductive at 10 V reading, as manifested by three peaks in the current profile.

Since the reading at the lower voltage of 3 V is performed immediately after the switching to LRS by the higher voltage of 10 V, the defect concentrations in the grains and grain boundaries are unlikely to change. Thus, both high and low external voltages probe the same SET state of the film. From these results, we can conclude that the grain boundaries at the SET state are two-dimensional conduction channels that are not activated at low reading voltages, but are sharply turned on at high reading voltages. The behavior of the grain boundaries is clearly that of a two-dimensional nanovaristor: they do not conduct until a critical voltage of about 10 V is applied, when they become highly conductive. If the voltage is subsequently decreased, the grain boundaries no longer conduct. This varistor behavior can be attributed to the onset of hopping conduction through a high- density planar distribution of oxygen vacancies $V_O$. The total film, however, goes through a hysteresis loop and a SET/RESET memristive switching.

We will now account for the complete memristive cycle by a detailed examination of multi-scale oxygen-vacancy dynamics aided by density functional theory (DFT) calculations and Monte-Carlo simulations. Density functional theory calculations are performed using the PBE version of the exchange-correlation functional [23], PAW potentials [24], and plane-wave basis as implemented in the VASP code [25]. Formation energies of oxygen vacancies at grain boundaries are calculated for the (111) grain surface using a 2x2x4 supercell with 160 atoms. Diffusion barriers are calculated using a 2x2x3 supercell with 120 atoms. We used 212 eV for the plane-wave cutoff and the Gamma-point for k-point sampling.

Oxygen-vacancy formation energies are already available in the literature [26]. At the relatively high annealing temperatures used in the sample fabrication (700°C), the Fermi energy tends to be in the mid-gap region, where the formation energy of the doubly positive oxygen vacancy is less than 1 eV. The result is typical for transition-metal oxides (TMOs) and consistent with the fact that most TMOs have large concentrations of oxygen vacancies [27-29]. DFT calculations were performed for oxygen-vacancy formation energies in grain boundaries and for diffusion energy barriers. Vacancy formation energies are typically ~0.3 eV smaller in grain boundaries, while diffusion barriers are 2.1-2.6 eV. As a consequence, during annealing at 700°C, the O vacancies diffuse and equilibrate in the grains and grain boundaries. At room temperature, however, thermal redistribution of the vacancies is essentially impossible so that the relative concentrations in the grains and grain boundaries remain unchanged. Though the vacancies in principle dope the material n-type, the energy levels are quite deep (0.6 and 1.6 eV for the first and second ionization of neutral vacancies; these values have large error bars [26, 30, 31]). Thus, the carrier concentration is quite low and the ionized vacancies act as Coulomb scattering centers, resulting in negligible electronic transport. This is the HRS of the material.

When an electric field is applied at room temperature, Poole-Frenkel emission [32, 33] occurs at oxygen vacancies (the electric field lowers the ionization energy). Subsequent non-radiative capture amounts to the capture energy (0.6 or 1.6 eV in the present cases) being converted to mostly localized phonons, known as a "phonon kick" which leads to what is known as recombination-enhanced migration [22]. This is the process that drives the non-equilibrium oxygen-vacancy concentrations to redistribute. It is natural for excess vacancies to be driven to the grain boundaries where they bind by an energy order of 0.3 eV.

When the external voltage is increasing, oxygen-vacancy agglomeration in the grain boundaries allows current to flow through the grains for several reasons: Poole-Frenkel emission contributes to the free carrier density and the concentration of Coulomb scattering centers is reduced in the grains (remote doping). The experimental data are consistent with this analysis: at 3V, the current is still effectively zero everywhere (~5 pA), but the grains conduct at 6 V (Fig. 1b,c). This current is probably space-charge limited [34]. The SET event, however, occurs at a large critical voltage, 10 V in this case (Fig. 1d and 2a), when the electric field in the grain boundaries is large enough to achieve a maximum concentration of

O vacancies *and* eliminate the energy barrier for electrons to hop from one oxygen vacancy to another. The elimination of the barriers is critical because, even at a high vacancy concentration, the energy levels vary due to the diverse local environment. The resulting hopping current in the grain boundaries constitutes the SET to a LRS (Fig. 1c suggests that some grain boundaries may carry some hopping current at voltages smaller than the SET voltage; this is possible if the environment of O sites happens to be more uniform than average and barriers for conduction may be smaller; see below for more detailed analysis of hopping conduction). We note that the conductive channels are not true metallic filaments as often suggested [11, 12]. The hopping current is turned on by a high voltage and the grain boundaries behave as true nanovaristors. Indeed, if the voltage is reduced as in Fig. 1e and Fig. 2b, conduction through the grain boundaries is cut off as hopping barriers re-appear, while conduction through the grains persists with larger current than prior to switching (Fig. 1b). Although the switched grains (bright parts in Fig. 1c-1e) are depleted of $V_O$ dopants, they are in fact more conductive than unswitched grains because the electron density remains essentially unchanged (the $V_O$'s that have been swept into the grain boundaries provide the carriers to the grains as in remote doping), while Coulomb scattering from $V_O$'s is reduced.

According to the Miller-Abrahams theory of hopping conduction [35], the rate $v$ of a carrier to hop from one localized state A to another localized state B (A and B are oxygen vacancies in this case) can be expressed as:

$$v_{A \to B} = v_0 \exp(-2r/r_0) \begin{cases} \exp(-\Delta E / kT), & \text{for } \Delta E > 0 \\ 1, & \text{for } \Delta E \leq 0 \end{cases}, \quad (1)$$

where $v_0$ is the attempt frequency for the hop, $r$ is the distance between the vacancies (the hopping distance), $r_0$ is the spatial extend of the localized state, and $\Delta E = E_B - E_A$ is the difference between the energy levels of the final and initial states.

A fraction of the connections between $V_O$'s have positive $\Delta E$ and can only be thermally activated, and as a result, the connecting cluster of vacancies in the grain boundary are not very conductive, as illustrated in Figure 3a(top). This corresponds to the experimentally observed lack of current through grain boundaries at relatively low voltages (Fig. 1e and Fig.

2b). In contrast, when a large electric field (voltage) is present between the electrodes, the value of $\Delta E$ changes to $E_B - E_A - \vec{E} \cdot \vec{r}$, where $\vec{E}$ is the electric field vector and $\vec{r}$ is the hopping displacement vector. At a sufficiently high electric field, the modification term $-\vec{E} \cdot \vec{r}$ for the near-neighbor connections along the field direction becomes comparable to the spread of the energy levels. Thus the positive $\Delta E$ becomes zero or negative, and thermal activation is no longer required. In this way, the connecting cluster becomes highly conductive, as illustrated in Figure 3a (bottom). This corresponds to the readings at high voltage in experiments, where a large current is observed at the boundaries.

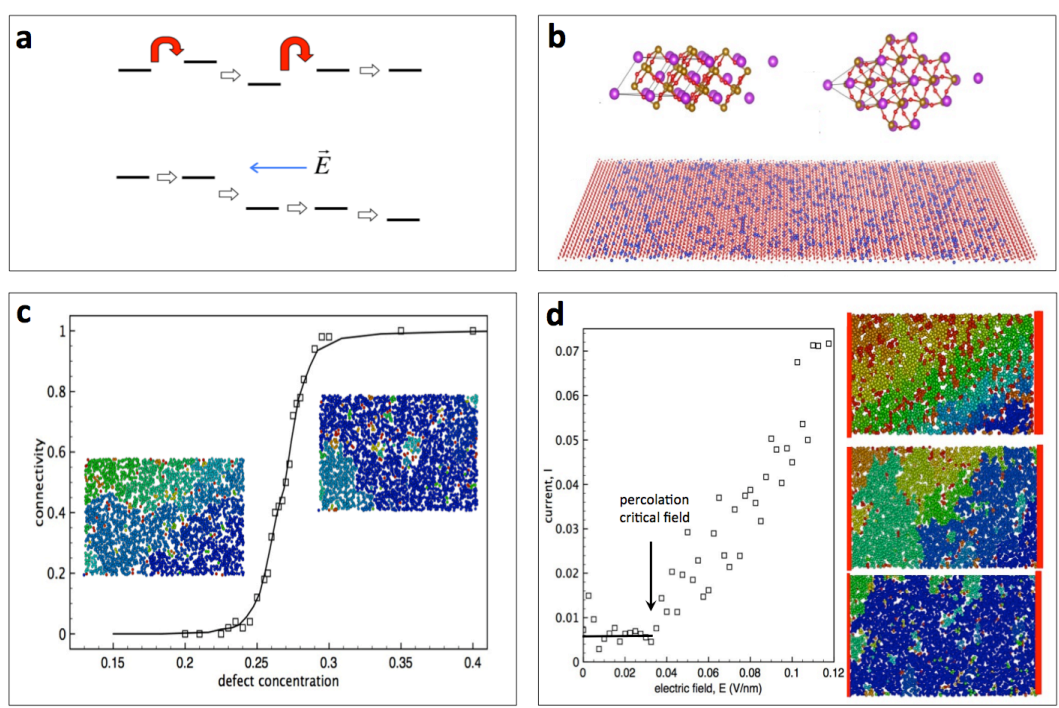

**Figure 3. Current through vacancies in grain boundaries. a**, Schematic of the energy profile of oxygen vacancies levels in a connecting cluster, projected along the direction of the applied electric field. At zero or low voltage, a fraction of the near-neighbor connections has positive $\Delta E$ and require additional activation (shown as red arrows in the top schematic). The connecting cluster is not very conductive. At the critical voltage, the barriers are eliminated and current flows (bottom schematic). **b**, Top schematics: side and top views of the atomic structure of BiFeO$_3$ at the grain surface, with Bi in purple, Fe in brown, and O in red. Bottom schematic: supercell with random arrangement of oxygen vacancies used in the percolation analysis and Monte Carlo transport calculation, with O sites in red and vacancies in blue. **c**, Connectivity dependence on oxygen vacancy concentration showing a threshold value of 0.25. Insets show connecting clusters at concentration below and above the threshold concentration, corresponding to non-spanning (left) and spanning (right) cases. Atomic configurations are color-coded according to the different connecting clusters they belong to. The dark blue is used to label the connecting cluster that is attached to the right electrode. **d**, Left: Current-field relation

calculated by Monte Carlo. The critical field to conduct is shown by the arrow. Right: Maps of conducting regions under different electrical field strengths. Atomic configurations are color-coded to represent the *conducting* regions they belong to. Within each conducting regions the electron transport (to the right) is allowed by *both* the *connectivity* and *energy levels*. The dark blue labels the conducting region through which electrons can reach the right electrode only above a critical voltage.

We performed a percolation analysis of the conducting region of a $BiFeO_3$ grain boundary with a random distribution of $V_O$ as we calculate the current-voltage relationship using the Monte Carlo method. The structure of $BiFeO_3$ at the grain surface is shown in Fig. 3b (top). The distribution of $V_O$ in the simulation supercell (10 nm by 10 nm for demonstration purposes) is shown in Fig. 3b (bottom). In the percolation analysis, we consider hopping within a range of 0.5 nm, which includes hopping between first and second nearest $V_O$ neighbors in BiFeO3. We chose 30% concentration of $V_O$ in our model calculation, as a spanning cluster that connects two leads exists only for concentrations above a threshold value, due to finite hopping radius. This is clear from Figure 3c, which shows the connectivity as a function of $V_O$ concentration and connecting cluster configurations at concentrations below and above the threshold value of 25% ("connectivity" in Fig. 3c is defined as the percent of vacancy sites that have at least one vacancy site as first or second neighbor). This analysis shows that a high concentration of vacancies at a grain boundary is essential for conduction. As the oxygen vacancies are initially distributed both at grain boundaries and within grains, larger grains are favorable for having memristive switching as they have large volume/surface ratios and thus more vacancies per surface area. This is consistent with a previous study showing that memristive switching in BFO only happens in large grains [18], although in that study the higher anneal temperature used to form the large grains may also have played a role in creating more vacancies.

As we already noted, oxygen vacancies in bulk $BiFeO_3$ are deep donors with charge states +1 and +2. The calculated values reported in the literature vary by 0.3 eV. We adopted the numbers 0.6 and 1.6 eV and performed Monte Carlo simulations of electron transport using Eq. (1) in the percolation model and obtained the current-field relations shown as Figure 3d (left). It is clear that there is a critical value of the electric field that allows conduction. Below the critical value, the current is limited despite the existence of the connecting cluster. This is due to the difference in the energy levels, leading to the separate conducting regions and shown in Fig. 3d (top- and center-right). The region that connects to the right lead grows with increase in the applied electrical field. As the field reached critical value, the size of this conducting region reaches the size of the entire supercell (Fig. 3d, bottom-right) and the current increases rapidly as shown in Fig. 3d (left).

The above percolation analysis and Monte Carlo simulations are in agreement with the experimental results of the local conduction at grain boundaries at LRS, confirming the validity of the conclusion that grain boundaries behave like nanovaristors. What is left now is to account for the RESET at -5 V, which converts the film to the initial state. We first note that, after SET, in the LRS, after steady is reached, the majority of vacancies are in the grain boundaries with the fully ionized vacancies in the 2+ charge state preferentially near the grounded electrode as shown schematically in Fig. 4a. They set up an electric field that opposes the external field, so that the effective SET field is considerably smaller (ionic screening), as shown in Fig. 4b. When a large negative electric field pulse (-5 V) is applied for RESET, the external and internal fields instantaneously add, before the charge on vacancies has a chance to redistribute. The net value of the field is much higher than the effective SET voltage (Fig. 4c) and produces local heating, i.e., it is effectively a heat pulse, that redistributes the vacancies back to the initial arrangement that was present in the field after the 700°C anneal. Although in an ideal situation the switching mechanism would allow SET with either positive or negative voltage, experimentally we only observe the SET with positive voltage. The preferred SET polarity may be caused by the different electrode geometries. In an ideal situation, both electrodes are flat and the electrical field is homogeneous. However, in our experiments, the bottom electrode (Pt substrate) is flat and the top electrode (AFM Pt tip) is sharp, which causes the distribution of electrical field to be inhomogeneous. The inhomogeneity of the electric field may be causing the preferred polarity in the SET.

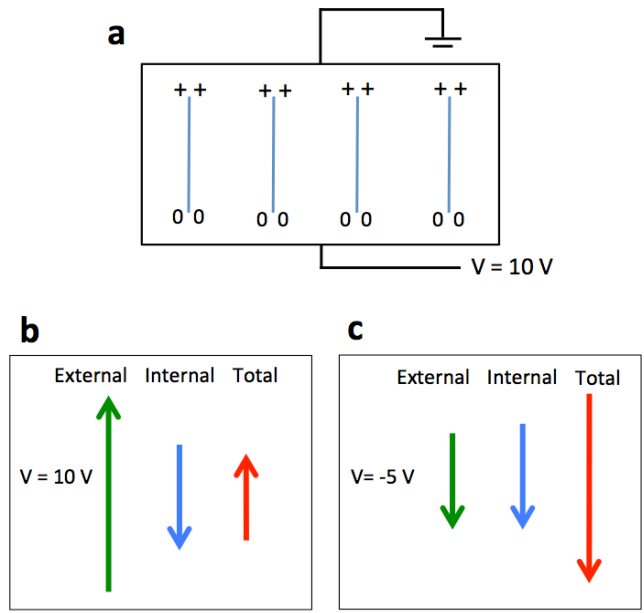

**Figure 4. Effective fields in a grain boundary with O vacancies. a**, Schematic diagrams of charge distribution of aggregated $V_O$ at grain boundaries after the steady state is reached at a 10 V SET voltage. **b**, Local field composition at the SET voltage. **c**, Local field composition at the moment a -5 V RESET voltage is applied.

In summary, we have given experimental evidence that, under an applied electric field, the grain boundaries in polycrystalline BFO behave like nanovaristors and described in detail that atomic-scale processes that account for the complete memristive cycle from SET to RESET. This behavior is likely to occur in other polycrystalline memristive materials. In addition, our Monte Carlo codes that reproduce the nanovaristor behavior can be used to parameterize and model real memristive materials for the design of memristive devices. The atomic-scale understanding provided by our analysis is likely to lead to novel memristive devices.

**Acknowledgments**

The theoretical work was supported by National Science Foundation grant DMR-1207241 and by the McMinn Endowment at Vanderbilt University. Computational support was provided by the NSF XSEDE under Grant # TG-DMR130121. The experimental work was supported by the National Basic Research Program of China (2011CB707601 and 2012CB933004), the National Natural Science Foundation of China under grant Nos. 11204034, 61274114, 113279028, and 11474295, the Natural Science Foundation of Jiangsu Province under grant Nos. BK2012123 and BK2012024, the Specialized Research Fund for the Doctoral Program of Higher Education of China (20120092120023) and Ningbo Science and Technology Innovation Team (2011B82004). In the USA, research was supported in part by the Department of Energy Office of Science, Basic Energy Sciences, Materials Science and Engineering Directorate (KY, XS), Department of Energy grant DE-FG02-09ER46554 (STP, YSP), and by the McMinn Endowment at Vanderbilt University (STP).